\documentclass[a4paper]{jpconf}
\usepackage{epsfig}
\usepackage{graphicx}
\usepackage{hyperref}
\begin{document}
\title{Mirage Cosmology on Unstable D3-Brane Universe}

\author{Anastasios Psinas\footnote{Present address: Department
of Physics, University of Patras, GR-26500, Patras, GREECE. Talk
presented at the RTN meeting ``The Quest for Unification: Theory
Confronts Experiment", Corfu, Greece, Sept, 11-18, 2005 }}

\address{Department of Physics, Northeastern University, Boston, MA 02115-5000, USA}

\ead{psinas.a@neu.edu}

\begin{abstract}
We study the geodesic motion of an unstable brane moving in a
higher dimensional bulk spacetime. The tachyon which is coupled to
a $U(1)$ gauge field induces a non-trivial cosmological evolution.
Interestingly enough, this system exhibits a much smoother initial
cosmological singularity in comparison with former works.
\end{abstract}
String cosmology has been one of the leading approaches towards
describing cosmos at large scale. Based on the very rich structure
of string theory and its dualities cosmologists have acquired a
better understanding of the evolution of the universe. In
particular, the progress made in the AdS/CFT \cite{1}
correspondence gave significant impetus towards the development of
various cosmological models. In one of those \cite{2}, the
universe is described by a D$p$-brane (with $p=3$ for a $4$-dim
universe) moving in a higher $p'$ dimensional space. According to
this, the ambient space is filled by a stack of $Dp'$-branes with
$p'>p$ which is much heavier than the probe so that the
back-reaction is neglected. The cosmological expansion or
contraction that these universes exhibit can be ascribed to mirage
matter that may come from the bulk. Such constructions have very
interesting consequences with respect to the problem of the
initial singularity. For instance, one can study the effects of
different kind of matter on the effective energy density of the
probe brane by investigating by how much and to what extent
cosmological singularities are regularized. Essentially, the
energy content of the theory can give us details of the spacetime
structure on the brane universe.

The instabilities in the spectrum of string theory (in the form of
tachyons) have been a serious issue for quite some time. Quite
recently though, it is shown, that unstable D-branes can form and
then decay so that we end up with a stable string vacuum
\cite{3,4}. It is therefore believed that the very early history
of the universe was dominated by instabilities of tachyonic
nature. Several implementations of the tachyon in cosmology are
known. In \cite{4} there has been an extensive analysis of mirage
cosmology for tachyons which live in the bulk. Models of unstable
universes are also obtained in the case of the geodesic motion of
a tachyonic D$p$-brane for which the tachyon field is confined on
the defect \cite{6}. Thinking in the same spirit, it would be
instructive to examine the possibility of having tachyon matter
coupled to an abelian gauge field where both of them reside on the
brane \cite{7}. Keep in mind that unlike the brane-antibrane
models \cite{8} we just consider one mobile D$3$-brane while the
gauge group is $U(1)$. We also expect the gauge degrees of freedom
to play an important role at early times. For instance, we recall
that even in the FRW model the state equation of the cosmic fluid
is that of a photonic gas during the first stages of the evolution
of cosmos. In fact, as it will be shown, our analysis clearly
points out that at very early times the universe is radiation
dominated.



We start with the most general ansatz of a ten dimensional static
and spherically symmetric spacetime

\begin{equation}
ds^2_{10}=g_{00}(r)dt^2+g(r)(d\overrightarrow{x})^2+
g_{rr}dr^2+g_{S}(r)d\Omega_{5} \label{1}
\end{equation}
where the timelike part of the metric $g_{00}$ is negative.
Additionally, the complex tachyon on the D$3$-brane couples to the
massless gauge field in the following fashion

\begin{equation}
S_{3}=-T_{3}\int{d^4\xi}e^{-\phi}V(T)\sqrt{-det\tilde{K}_{\mu\nu}}-T_{3}\int{d^4\xi}\hat{C_{4}}\label{2}
\end{equation}
where
\begin{equation}
\tilde{K}_{\mu\nu}=K_{MN}\frac{\partial x^{M}}{\partial
x^{\mu}}\frac{\partial x^{N}}{\partial x^{\nu}} \label{3}
\end{equation}
and
\begin{equation}
K_{MN}=g_{MN}+2\pi{\alpha}'
F_{MN}+\frac{1}{2}(D_{M}TD_{N}T^{*}+c.c) \label{4}
\end{equation}
We ascribe capital indices M,N for the bulk coordinates while
$\mu,\nu$ are the one kept for the probe brane. We note that
$F_{\mu\nu}$ is the world volume antisymmetric gauge field
strength and $g_{MN}$ is the induced metric on the D$3$-brane
universe and $ \phi $ is the dilaton. Observe, that the RR field
$\hat{C}(r)=C_{0...3}(r)$ is also included in the action. In
addition, the form of the complex tachyon field and its covariant
derivative are given as follows
\begin{equation}
D_{M}T=\partial_{M} T - iA_{M}T \label{5}
\end{equation}
\begin{equation}
T=T_{x}+iT_{y} \label{6}
\end{equation}
Here we underline that for the rest of the analysis we will
consider that $dV(T)/dT\simeq{0}$. This is equivalent to say that
we are in the regime of small $T$ . In other words we are mainly
focused on exploring the dynamics of the D-brane at a point which
the instability has started to occur.

In our analysis we assign an expectation value to the tachyon,
i.e. $T_y=c$ while we adopt the gauge $ A_{0}=0$. The remaining
non-vanishing spatial components $A_{i}$ define the following
combination

\begin{equation}
\alpha^{i}=A^{i}_{0}c \label{10}
\end{equation}
It is crucial to mention that we study small and time-dependent
fluctuations of the gauge field around $A^{i}_{0}$ which is taken
to be small. All of the above translate to

\begin{equation}
A^{i}(t)= A^{i}_{0}+ \tilde{A}^{i}(t) \label{8}
\end{equation}
Thus our Lagrangian after some obvious redefinitions turns out to
be

\begin{equation}
L=-\sqrt{M-B\dot r^2-Ch_{ij}\dot \varphi^{i}\dot \varphi^{j}-D\dot
T^2-E(\alpha^{i}\dot A_{i}\dot T+ \dot A_{i}\dot A^{i})}+
\hat{C}(r) \label{11}
\end{equation}

The presence of the radical in the lagrangian will certainly
complicate the form of the equation of motions of the associated
fields \cite{7}. Expanding the square root is a good approximation
only at late times since for small scale factors $a(r(t))$ the
above approximation scheme fails to capture the underlying physics
at the $a\rightarrow{0}$ limit. Besides, the tachyon induced
instabilities will be more prominent as $a\rightarrow{0}$ since at
very late times the tachyon will condense. In addition, it would
be interesting to examine the above system for small scale factors
so that we improve our understanding off the tachyon-gauge
field-gravity interactions at high curvature.

 Interestingly enough, the induced metric on the brane
 describes a flat FRW spacetime which reads

\begin{equation}
d\hat s^2=-d\eta^2+g(r(\eta))(d\overrightarrow{x})^2 \label{29}
\end{equation}
while the cosmic time is defined as

\begin{equation}
d\eta=\frac{|g_{00}|}{\mathcal{E}}\sqrt{T^2_{3}V^2_{0}e^{-2\phi}g^3
+Q_{i}Q_{i}
+\frac{(2g\tilde{Q}-\alpha_{i}Q_{i})^2}{(4g^{2}-\alpha_{i}
\alpha_{i})}}dt \label{30}
\end{equation}
For the sake of clarity we note that $\mathcal{E}$ denotes the
energy of the system while $Q_{i}$ and $\tilde{Q}$ stand for the
integration constants of the gauge field and the tachyon
respectively. Finally, due to the spherical symmetry of the
ambient space the angular momentum $l$ is conserved.

Close inspection of Eq.~(\ref{30}) reveals that the argument of
the radical is not only non-positive but it is also singular as
well when $ 4g^2(r)=\alpha_{i}\alpha_{i}$ unlike in models like
\cite{2}. Nonetheless, we are primarily interested in the regime
of very small scale factors. At that limit, it is easy to check,
that the argument in the radical is strictly nonnegative (for
nonzero values of the $Q_{i}$ charges) while the singularity has
been avoided. Our analysis so far pertains a very general class of
metrics (equipped with certain symmetries) describing the bulk. A
very characteristic example of a spacetime of this short, is the
well studied case of the near-horizon limit of a black hole in an
$AdS_{5}\times{S^{5}}$ background

\begin{equation}
ds^2=\frac{r^2}{L^2}(-f(r)dt^2+(d\overrightarrow{x}^2))
+\frac{L^2}{r^2}\frac{dr^2}{f(r)}+L^2d\Omega^2_{5} \label{35}
\end{equation}
where $f(r)=1-\frac{r_{0}^4}{r^4}$ and $\hat{C}(r)=
\frac{r^4}{L^4}-\frac{r_{0}^4}{2L^4}$\\
Given the very simple form of the induced metric on the brane, one
can straightforwardly derive the FRW equations. Thus, the
effective energy density on the probe reads as follows

\begin{equation}
\frac{8\pi G}{3}\rho_{\mathrm{eff}}=
\frac{1}{a^2L^2F^2}((\mathcal{E}+a^4)^2-a^2(1-\frac{r_{0}^4}{L^4a^4})
(F^2+\frac{l_{S}^2}{L^2})) \label{36}
\end{equation}
The constant term accompanying the RR term is absorbed in
$\mathcal{E}$. Also, F is given as a function of the scale factor
of the universe in the following form

\begin{equation}
F^2=T^2_{3}V_{0}^2e^{-2\phi}a^6+Q_{i}Q_{i}
+\frac{(2\tilde{Q}a^2-\alpha_{i}Q_{i})^2}{4a^4-\alpha_i\alpha_{i}}
\label{37}
\end{equation}

In principle, the singularities a given spacetime exhibits can be
related to the energy content of the theory living in a given
background. Thus, it is reasonable to expect that the blowing up
of the energy density on the brane is indicative of a curvature
(initial) singularity. Thus, we are interested in exploring the
behavior of $\rho_{\mathrm{eff}}$ as $a\rightarrow{0}$.
\footnote[2]{An alternative method of characterizing a singularity
as a curvature singularity is through the evaluation of the
Kretschmann scalar
$R_{\mu\nu\kappa\lambda}R^{\mu\nu\kappa\lambda}$. However in the
context of our analysis this method is not preferable.}

Let us now study in detail the behavior of the energy density in
the limit of very small scale factors. It turns out, that as
$a\rightarrow{0}$ $F^2$ reads as follows

\begin{equation}
F^2=T_{3}^2V_{0}e^{-2\phi}a^6+|\overrightarrow{Q}|^2\sin^2{\theta}
\label{40}
\end{equation}
where $\theta$ denotes the angle between vectors $\alpha_{i}$ and
$Q_{i}$. One observes that $\rho_{\mathrm{eff}}$ has no explicit
dependance on $\alpha_{i}$ which manifests the non trivial
interaction between the tachyon and the gauge field. However, the
interaction appears implicitly through the
$|\overrightarrow{Q}|^2\sin^2{\theta}$ term. Also, note that
whenever $\sin{\theta}$ becomes zero the effective energy density
on the unstable probe may be similar to the one of a stable
D$3$-brane\cite{2}. In all other cases this novel effect doesn't
occur due to the non vanishing integration constants. We also
stress, that the singular behavior that the $F^2$ term exhibited
in the most general case, has disappeared at the very early stages
of the evolution of the universe.

A simpler version of Eq.~(\ref{36}) is obtained when the gauge
field vanishes. It turns out that in this case the presence of
tachyon regulates the cosmological evolution by smoothening the
initial singularity. In essence, our analysis corroborates former
results found in \cite{6} for a vanishing Ramond-Ramond field.

For the shake of concreteness we denote the full expressions of
the energy density and effective pressure on the brane universe in
the $a\rightarrow{0}$ limit. To this end, we express
$\rho_{\mathrm{eff}}$ and $p_{\mathrm{eff}}$ in powers of inverse
scale factor so that it is easier to read the value of $w$ which
enters in the state equation
$p_{\mathrm{eff}}=w\rho_{\mathrm{eff}}$ of the cosmic fluid. Thus,
we obtain


\begin{eqnarray}
 p_{\mathrm{eff}}= \frac{1}{8\pi GL^2}\frac{r_{0}^4}{L^4}
 (1+\frac{l^2}{L^2}\frac{1}{Q_{i}^2\sin^2{\theta}})\frac{1}{a^4}\nonumber\\
+(\frac{\mathcal{E}^2}{8\pi
 GL^2})(-\frac{1}{(Q_{i}^2\sin^2{\theta})^2}
+\frac{8r_{0}^4}{L^4}\frac{l^2}{L^2}\frac{\alpha_{i}Q_{i}\tilde{Q}}
{\alpha_{i}^2(Q_{i}^2\sin^2{\theta})^2})\frac{1}{a^2} \nonumber \\
+\frac{3}{8\pi
 GL^2}(1+\frac{l^2}{L^2}\frac{1}{Q_{i}^2\sin^2{\theta}}
+\frac{8\mathcal{E}^2}{3}
\frac{\alpha_{i}Q_{i}\tilde{Q}}{\alpha_{i}^2(Q_{i}^2\sin^2{\theta})^2}
-\frac{16}{3}\frac{r_{0}^4}{L^4}\frac{l^2}{L^2}\frac{(\alpha_{i}Q_{i})^2}{\alpha_{i}^4})
\label{43}
\end{eqnarray}
while the effective density is given by

\begin{equation}
\rho_{\mathrm{eff}}=\frac{3}{8\pi
GL^2}\frac{r_{0}^4}{L^4}(1+\frac{l^2}{L^2}\frac{1}{Q_{i}^2\sin^2{\theta}})\frac{1}{a^4}
+\frac{3\mathcal{E}^2}{8\pi
GL^2}\frac{1}{(Q_{i}^2\sin^2{\theta})^2}\frac{1}{a^2}-
\frac{3}{8\pi
GL^2}(1+\frac{l^2}{L^2}\frac{1}{Q_{i}^2\sin^2{\theta}}) \label{44}
\end{equation}

At this point, let us comment on Eq.~(\ref{43},\ref{44}) in
detail. First of all, it is clear that the energy density has no
explicit dependance on the tachyon integration constant
$\tilde{Q}$ while only the gauge field constants $Q_{i}$ survive.
It looks as if the tachyonic degrees of freedom have no effect on
$\rho_{\mathrm{eff}}$ as $a\rightarrow{0}$ when a gauge
field-tachyon interaction is present. However, this is not the
case with the effective pressure on the brane. In addition, it
turns out that $w=1/3$ at the very early history of the universe
leading one to the conclusion that the universe is radiation
dominated. Thus, initially the gauge field prevails over the
tachyon. At later evolutionary stages $w$ depends on the values of
the parameters of the system ($Q_i$, $\tilde{Q}$, $\alpha_{i}$,
$\sin{\theta}$, $\mathcal{E}$, $l$, $r_0$, $L$) and so causality
may be violated. Closer inspection of Eq.~(\ref{43},\ref{44})
reveals that at late times a fine tuning of the parameters can
lead to a pressureless universe with a finite energy density which
could be negative. The negativity of the energy density is a
manifestation of the tachyon.

 An equally interesting case is the one with a vanishing
 $\alpha_{i}$. One can easily show that $w$ takes the following
 values $w=[\frac{1}{3},-\frac{1}{3},-1]$. Initially, the probe is
 radiation dominated however at later times it has the state of a
 universe filled with a cosmic fluid that resembles a
 non-vanishing cosmological constant. So, even though
 $w=-1$ the associated energy density is negative due to the
 tachyon. Interestingly enough, in this case, $|w|\leq{1}$ and
 thus causality is respected. The above values for $w$ are reminiscent of
quintessence in which the accelerated expansion of the universe
occurs when $-1<w<-\frac{1}{3}$ \cite{9}. For completeness, the
term $\sim{a^{-2}}$
 in $\rho_{\mathrm{eff}}$ has the same effect as a negative curvature
 term responsible for accelerating the expansion of the universe.

The effects of the minimal tachyon-gauged boson coupling on the
initial cosmological singularity are quite transparent. According
to Eq.~(\ref{44}) the energy density is not that divergent as
$a\rightarrow{0}$ in comparison to former works \cite{2}. More
specifically, in the model at hand, the most divergent term of the
energy density goes like $\sim{a^{-4}}$ while for stable D-branes
\cite{2} it is shown that $\rho_{\mathrm{eff}}\sim{a^{-8}}$. This
in particular leads one to the conclusion that the above coupling
makes the singularity of spacetime less severe. For purely
tachyonic branes a similar result was initially reported in
\cite{6}.



In our approach we were mainly interested in dealing with unstable
D$3$-branes moving in a geodesic embedded in a higher dimensional
bulk. Although in models like ours it is very hard to obtain
solutions in closed form we didn't face such a problem. This can
be attributed to the fact that we were focused in a region where
the tachyon potential was very flat. The presence of the minimal
gauge field-tachyon coupling was shown to have significant
cosmological consequences. According to our analysis, the universe
is proven to be radiation dominated at its incipient stages while
at later epochs it can evolve in a more complicated fashion. This
behavior clearly shows that the gauge degrees of freedom are
prevalent, however as the universe evolves the effects of the
tachyon become more apparent. Further, it is shown, that the
initial singularity which the mirage models exhibit is softened
quite significantly. This particular result looks quite promising
even though we were not able to completely remove the singularity.
We hope that further work in this direction will provide more
clues on how to tame cosmological singularities.\\

{\bf Acknowledgements}\\

I would like to thank the organizers of the RTN meetings ``The
Quest for Unification: Theory Confronts Experiment" and
``Constituents, Fundamental Forces, and Symmetries of the
Universe" held in Corfu, Greece, September 11-26, 2005, for
financial support and for providing me the opportunity to present
the results of my work.




\end{document}